\documentclass{article}
\usepackage{ijcai22}

\usepackage{times}
\usepackage{soul}
\usepackage{url}
\usepackage[hidelinks]{hyperref}
\usepackage[utf8]{inputenc}
\usepackage[small]{caption}
\usepackage{graphicx}
\usepackage{amsmath}
\usepackage{amsthm}
\usepackage{algorithm}
\usepackage{algorithmic}
\usepackage{multirow}
\usepackage{tabularx}
\newtheorem{problem}{Problem}
\newtheorem{definition}{Definition}[section]
\usepackage{balance}
\newcolumntype{P}[1]{>{\centering\arraybackslash}p{#1}}
\usepackage{arydshln}
\usepackage{booktabs,siunitx}
\usepackage{bm}
\usepackage{pifont}
\usepackage{balance}
\usepackage{amsfonts}
\usepackage{xcolor,soul,lipsum}
\usepackage{nicematrix}

\usepackage[T1]{fontenc}
\usepackage[utf8]{inputenc}
\usepackage{babel}
\usepackage[font=small,labelfont=bf]{caption}

\usepackage{graphicx}
\usepackage{capt-of}
\usepackage{booktabs}
\usepackage{varwidth}
\newsavebox\tmpbox
\DeclareCaptionLabelFormat{andfigure}{#1~#2  \&  \figurename~\thefigure}

\makeatletter
\def\adl@drawiv#1#2#3{%
        \hskip.5\tabcolsep
        \xleaders#3{#2.5\@tempdimb #1{1}#2.5\@tempdimb}%
                #2\z@ plus1fil minus1fil\relax
        \hskip.5\tabcolsep}
\newcommand{\cdashlineCustom}[1]{%
  \noalign{\vskip\aboverulesep
           \global\let\@dashdrawstore\adl@draw
           \global\let\adl@draw\adl@drawiv}
  \cdashline{#1}
  \noalign{\global\let\adl@draw\@dashdrawstore
           \vskip\belowrulesep}}
\makeatother

\title{RecipeRec: A Heterogeneous Graph Learning Model for Recipe Recommendation}

\author{
Yijun Tian$^1$,
Chuxu Zhang$^2$,
Zhichun Guo$^1$,
Chao Huang$^3$,
Ronald Metoyer$^1$,
Nitesh V. Chawla$^1$\\
\affiliations
$^1$Department of Computer Science, University of Notre Dame, USA\\ 
$^2$Department of Computer Science, Brandeis University, USA\\
$^3$Department of Computer Science, University of Hong Kong, Hong Kong\\
\emails
$^1$\{yijun.tian, zguo5, rmetoyer, nchawla\}@nd.edu,
$^2$chuxuzhang@brandeis.edu,
$^3$chaohuang75@gmail.com
}

\begin{document}
\maketitle

\begin{abstract}
Recipe recommendation systems play an essential role in helping people decide what to eat. Existing recipe recommendation systems typically focused on content-based or collaborative filtering approaches, ignoring the higher-order collaborative signal such as relational structure information among users, recipes and food items. In this paper, we formalize the problem of \textit{recipe recommendation with graphs} to incorporate the collaborative signal into recipe recommendation through graph modeling. In particular, we first present \textit{URI-Graph}, a new and large-scale user-recipe-ingredient graph. We then propose \textit{RecipeRec}, a novel heterogeneous graph learning model for recipe recommendation. The proposed model can capture recipe content and collaborative signal through a heterogeneous graph neural network with hierarchical attention and an ingredient set transformer. We also introduce a graph contrastive augmentation strategy to extract informative graph knowledge in a self-supervised manner. Finally, we design a joint objective function of recommendation and contrastive learning to optimize the model. Extensive experiments demonstrate that \textit{RecipeRec} outperforms state-of-the-art methods for recipe recommendation. Dataset and codes are available at \url{https://github.com/meettyj/RecipeRec}.

\end{abstract}

\section{Introduction}

Large-scale food data offers rich knowledge about the food landscape and can help address many important issues of today's society. Recipe websites are a popular source of food inspiration and enable users to rate and comment on recipes as well as receive recipe recommendation \cite{frontiers_recipe_rec}. In particular, \textit{www.food.com}, one of the largest recipe-sharing websites in the world, consists of over half a million recipes -- reflecting the popularity of online recipe portals and the great demand for recipe recommendation services. However, digging into this overwhelming amount of online recipe resources to find a satisfying recipe is very difficult \cite{hard_to_find_satisfying_recipe}. Thus, recipe recommendation systems are proposed to help users identify recipes that align with users' preferences and needs among numerous online resources \cite{ensemble_modeling}.

\begin{figure}
	\centering
	\includegraphics[width=0.9\columnwidth]{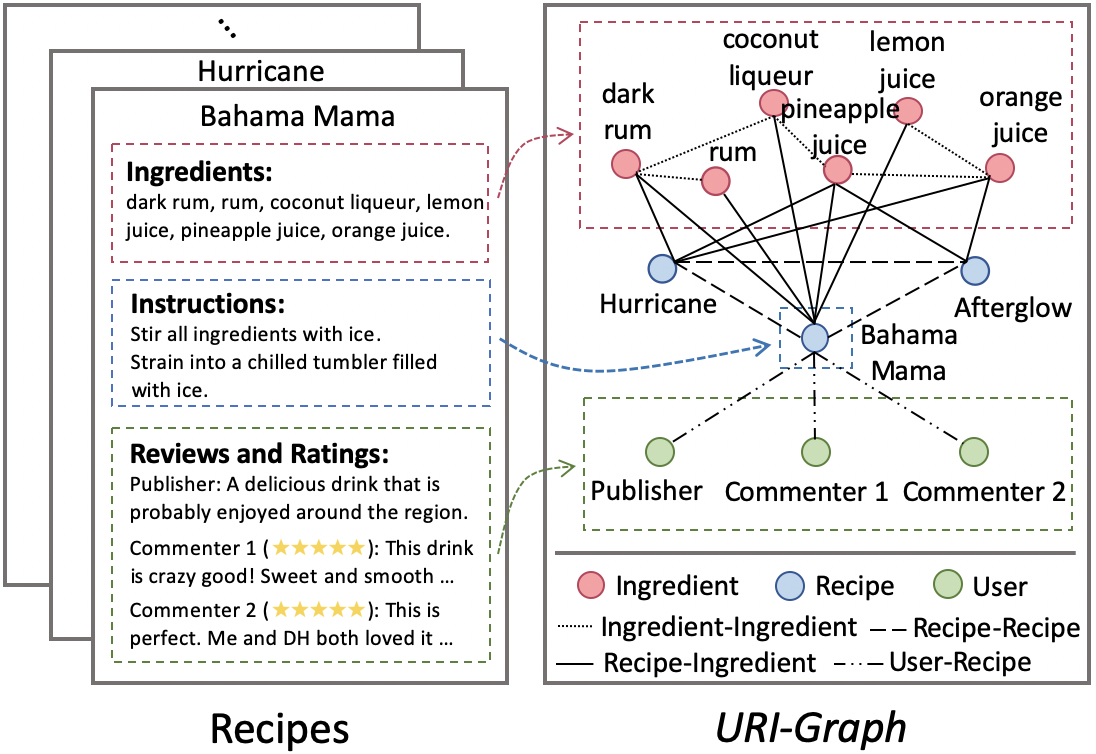}
	\caption{Illustration of \textit{URI-Graph}. Recipe examples are shown on the left. The \textit{URI-Graph} includes three types of nodes and four types of relations that connect these nodes.}
	\label{fig:fig1}
\end{figure}

Existing recipe recommendation approaches typically rely on computing similarities between recipes \cite{eating_healthier}. A few approaches attempt to take user information into account \cite{ensemble_modeling,visually_aware_rec}, but they only define similar users based on the overlapping rated recipes between users. This overlooks the higher-order collaborative signal (e.g., relational structure information) \cite{ngcf} present among ingredients, recipes, and users. However, individual preference for foods is complex. A user may decide to try a new recipe, for example, because of its ingredients, its cuisine type, or perhaps based on a friend's recommendation. Thus, a comprehensive recipe recommendation system should consider all of these factors. In particular, it is important to encode the higher-order collaborative signal into recipe recommendation and model the relationship among ingredients, recipes and users.

Therefore, we propose the problem of \textit{recipe recommendation with graphs}, which leverages the advances of graph modeling to address recipe recommendation. With the higher-order collaborative signal encoded by a graph learning model, the recommendation system can successfully capture the hidden collaborative filtering effect. Due to the complex relationship among ingredients, recipes and users (e.g., complementary ingredients, similar recipes), solving recipe recommendation with graphs requires an extensive and comprehensive recipe graph. Hence, it should not be surprising that the lack of research on this topic could be due to the lack of such a dataset. With the aim of making better recipe recommendation, we create \textit{URI-Graph}, a new and large-scale user-recipe-ingredient graph for recipe recommendation. As shown in Fig.\ref{fig:fig1}, we extract ingredients, instructions, and user interactions of each recipe to build \textit{URI-Graph}, which consists of three types of nodes and four types of relations.

Based on the created \textit{URI-Graph}, we propose \textit{RecipeRec}, a novel heterogeneous graph learning model for recipe recommendation. In particular, we first design a heterogeneous GNN to capture both recipe content and the relational structure information in \textit{URI-Graph}. To fully encode this information, we propose an adaptive node-level attention module to distinguish the subtle difference of neighboring nodes under a specific relation. We also introduce a relation-level attention module to select the most meaningful relations surrounding a node and automatically fuse them with proper weights. Then, we apply a set transformer~\cite{set_transformer} to encode the ingredient information and model the pairwise interactions and higher-order interactions among ingredients. Next, we introduce a graph contrastive augmentation strategy to extract informative graph knowledge in a self-supervised manner. Finally, we design a novel combined objective function of recommendation and graph contrastive learning to optimize the model. 
To summarize, our major contributions in this paper are as follows:
\begin{itemize}
  \item To the best of our knowledge, this is the first attempt to study \textit{recipe recommendation with graphs}. We create \textit{URI-Graph}, a new and large-scale recipe graph dataset, that can facilitate  recipe recommendation research and graph-based food studies.

  \item We propose \textit{RecipeRec}, a novel heterogeneous graph learning model for recipe recommendation. \textit{RecipeRec} can capture recipe content and relational structure information through various modules, including GNN with hierarchical attention, ingredient set transformer and graph contrastive augmentation. We also design a novel objective function to optimize the model.

  \item We conduct extensive experiments to evaluate the performance of our model. The results demonstrate the superiority of \textit{RecipeRec} by comparing it with state-of-the-art baselines for recipe recommendation.
\end{itemize}

\begin{figure*}[t]
	\centering
	\includegraphics[width=\textwidth]{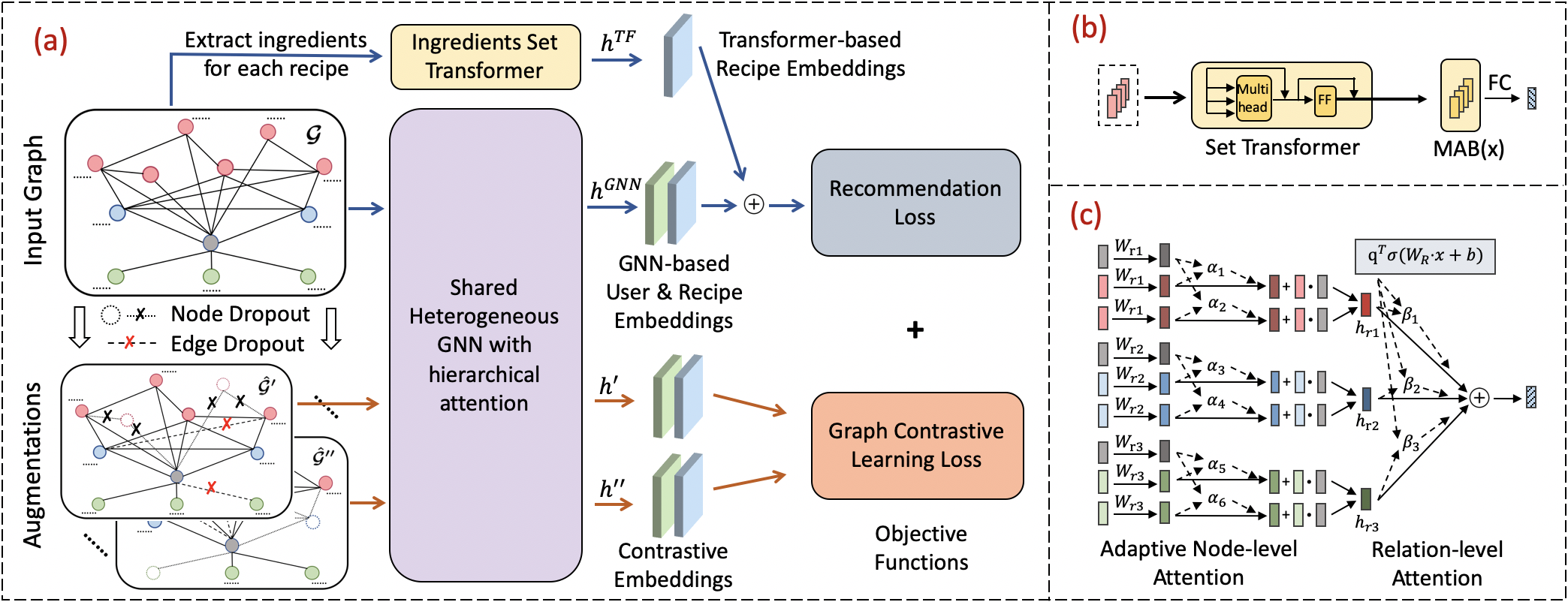}
	\caption{
	(a) The overall framework of our model: we first propose a heterogeneous GNN with hierarchical attention and an ingredient set transformer to learn user and recipe embeddings, which are later fed into the recipe recommendation loss. We then introduce a graph contrastive augmentation strategy to extract informative graph knowledge in a self-supervised manner. Next, the joint loss of recipe recommendation and graph contrastive learning is used to optimize the model;
	(b) Ingredient set transformer: encoding the interactions among ingredients with permutation invariant preserved; 
	(c) Heterogeneous GNN with hierarchical attention: encoding the nodes of the same type through adaptive node-level attention and then fusing the embeddings over different relations through relation-level attention.
	}
	\label{fig:fig2}
\end{figure*}

\section{Preliminary}
In this section, we describe the concept of user-recipe-ingredient graph, and formally define the problem of \textit{recipe recommendation with graphs}. We further introduce the \textit{URI-Graph} data that we create and use in this paper.

\begin{definition}
{\bf User-Recipe-Ingredient Graph.}
A user-recipe-ingredient graph is defined as a heterogeneous graph $\mathcal{G}=(V,E,C)$ with three types of nodes $V$ (i.e., user, recipe, ingredient) and four types of edges $E$ (i.e., ingredient-ingredient, recipe-ingredient, recipe-recipe, and user-recipe relations). In addition, nodes are associated with contents $C$, (e.g., instructions, nutrients, or ratings).
\end{definition}

\begin{problem}
{\bf Recipe Recommendation with Graphs.} Given a user-recipe-ingredient graph $\mathcal{G} = (V, E, C)$, the task is to design a machine learning model $\mathcal{F}_\Theta$ with parameters $\Theta$ to recommend recipes to users. In particular, the task is to return a ranked list of recipes for each user, such that its top ranked recipes are true favorite recipes of users. The recommender is supposed to take advantage of the associated recipe contents $C$ and relational structure information in $\mathcal{G}$.
\end{problem}

\noindent
\textbf{URI-Graph Data.}
To solve the problem, we first build \textit{URI-Graph}, as shown in Fig.\ref{fig:fig1}. We start by collecting recipes from Reciptor \cite{Reciptor} and crawl the user ratings for each recipe from \textit{food.com} as the user-recipe interactions. We then match each ingredient to the USDA nutritional database \cite{usda_sr} to get the nutritional information. Next, we build \textit{URI-Graph} by representing the users, recipes and ingredients as nodes, and constructing four types of relations to connect them. Specifically, we first connect each recipe and its ingredients with an edge, denoting the recipe-ingredient relation, while the weight of each ingredient is used as the edge weight. We then connect recipe nodes by the similarity determined by FoodKG \cite{foodkg} and the score is used as the edge weight, as shown in Reciptor \cite{Reciptor}. We further connect ingredient nodes by the co-occurring probabilities using Normalized Pointwise Mutual Information \cite{rn2vec}. We also connect users and recipes based on the interactions, with the rating scores as the edge weight. In the end, we build the graph with 7,958 user nodes, 68,794 recipe nodes and 8,847 ingredient nodes. Moreover, the graph contains four types of edges, including 135,353 user-recipe, 647,146 recipe-recipe, 463,485 recipe-ingredient and 146,188 ingredient-ingredient relations.

\section{RecipeRec}
To address recipe recommendation over the graph, we develop \textit{RecipeRec}, a heterogeneous graph learning model for recipe recommendation. As illustrated in Fig.\ref{fig:fig2}, \textit{RecipeRec} contains a heterogeneous GNN with hierarchical attention and an ingredient set transformer, with a graph contrastive augmentation strategy and a joint objective function.

\subsection{GNN with Hierarchical Attention}
\label{Section_GNN}
We first develop a heterogeneous GNN with hierarchical attention to obtain effective user and recipe embeddings. Specifically, for a node $v_i \in \mathcal{G}$, we use a type-specific input projection to transform the input feature, and leverage the adaptive node-level attention module to obtain a relation-specific embedding $h_{i,r}$ for each relation $r$ that connects to $v_i$. Then, we propose the relation-level attention module to fuse all relation-specific embeddings and obtain the final GNN-based node embedding $h_{i}^{G}$.

\noindent
\textbf{Type-specific Input Projection.}
For the input features of each node type, we use nutrient vectors to represent the ingredients ($x_{ing}$), the average of the pretrained skip-instruction embeddings \cite{recipe1m_1} to represent the recipes ($x_{ins}$) and random initialized vectors to represent the users ($x_{user}$). Due to the heterogeneity of the associated information, given a node $v_i$ of type $\phi_i$, we introduce a type-specific input projection $W_{\phi_i} \in \mathbb{R}^{{d_{\phi_i}} \times d}$ to project the input features into a shared embedding space:
\begin{equation}
\begin{aligned}
x_i &=
\begin{cases}
x_{ing}, & \text{if $\phi_i$ = ingredient}\\
x_{ins},& \text{if $\phi_i$ = recipe}\\
x_{user}, & \text{if $\phi_i$ = user}\\
\end{cases}\\
h_i &= W_{\phi_i} \cdot x_i,
\end{aligned}
\end{equation}
where $x_i$ is the input feature of $v_i$ with dimension $d_{\phi_i}$, and $h_i$ is the projected feature of $v_i$ with dimension $d$.

\noindent
\textbf{Adaptive Node-level Attention.}
To compute the relation-specific embedding $h_{i,r}^{l+1}$ for $v_i$ in layer $l+1$ of GNN, we first apply a shared weight matrix $W_r^l \in \mathbb{R}^{{d_l} \times d_{l+1}}$ to transform the input features from layer $l$, where $d_l$ and $d_{l+1}$ are the dimensions of embeddings in layer $l$, $l+1$ respectively:
\begin{equation}
z_{i}^l=W_r^l \cdot h_{i}^l,
\end{equation}
where $z_{i}^l \in \mathbb{R}^{d_{l+1}}$ is the intermediary embedding of $v_i$. We then calculate the unnormalized attention score $e_{ij}^l$ between $v_i$ and $v_j$, and normalize the $e_{ij}^l$ using $softmax$ function:
\begin{equation}
\begin{aligned}
e_{ij}^l &= \text{LeakyReLU}\left[W_{ij}^l \cdot (z_{i}^l\|z_{j}^l)\right],\\
\alpha_{ij}^l &= \frac{\exp(e_{ij}^l)}{\sum_{k\in N_{i,r} }^{}\exp(e_{ik}^l)},
\end{aligned}
\end{equation}
where $\|$ is concatenation, $W_{ij}^l\in \mathbb{R}^{2d_{l+1} \times d_{l+1}}$ is a weight vector, $N_{i,r}$ denotes the index set of neighboring nodes that connect to $v_i$ through relation $r$ and $\alpha_{ij}^l$ is the attention weight between $v_i$ and $v_j$. We use $\alpha_{ij}^l$ as coefficients to linearly combine the neighboring nodes features. We further make it adaptive to recommendation task by incorporating the interaction between $v_i$ and $v_j$ so that the interaction between users and recipes can be formulated attentively \cite{ngcf}. We also extend it to multi-head attention to make the training process stable. The process is formulated as follows:
\begin{equation}
h_{i,r}^{l+1}=
\overset{M}{\underset{m=1}{\Vert}} \text{ReLU} \left(\sum_{j\in N_{i,r}} {\alpha_{ij}^l \cdot z_{j}^l + W_h^l \cdot (h_i^l \odot h_j^l)}\right) \cdot W_a,
\end{equation}
where $h_{i,r}^{l+1}$ is the encoded embedding of layer $l+1$ for $v_i$ through relation $r$, $\odot$ denotes the element-wise product, $W_h^l \in \mathbb{R}^{{d_l} \times d_{l+1}}$ and $W_a \in \mathbb{R}^{Md_{m} \times d_{l+1}}$ are trainable weights, $M$ is the number of attention heads, and $d_{m}$ is the dimension of attention heads which satisfies $d_{m} = d_{l+1}/M$.

\noindent
\textbf{Relation-level Attention.}
We further introduce the relation-level attention to learn the importance of each relation and fuse all relation-specific node embeddings. In particular, we first apply a shared non-linear weight matrix $W_R \in \mathbb{R}^{d_{l+1}\times d_{l+1}}$ to transform the relation-specific node embeddings. Then we use a vector $q \in \mathbb{R}^{d_{l+1}}$ and calculate the similarities and average it for all node embeddings of relation $r$ to obtain the importance score $w_{i,r}$ for node $v_i$:
\begin{equation}
w_{i,r} =\frac{1}{|V_r|}\sum_{i \in V_r} q^\text{T} \cdot \tanh(W_{R} \cdot h_{i,r}^{l+1}+b),
\end{equation}
where $V_r$ denotes the set of nodes that are correlated with relation $r$ and $b \in \mathbb{R}^{d_{l+1}}$ is the bias vector. We then normalize the $w_{i,r}$ to get the final relation-level attention weight $\beta_{i,r}$:
\begin{equation}
\beta_{i,r}=\frac{\exp(w_{i,r})}{\sum_{r \in R_i} \exp(w_{i,r})},
\end{equation}
where $R_i$ indicates the associated relations of $v_i$ and $\beta_{i,r}$ can be explained as the contribution of relation $r$ to $v_i$. After that, we fuse the relation-specific node embeddings $h_{i,r}^{l+1}$ with $\beta_{i,r}$ to obtain the node embedding $h_i^{l+1}$:
\begin{equation}
h_{i}^{l+1}=\sum_{r=1}^{R_i} \beta_{i,r} \cdot h_{i,r}^{l+1}.
\label{eq_final_embedding}
\end{equation}
We denote the embedding obtained in the last propagation layer as the final GNN-based node embedding $h_i^{G}$ such that $h_i^{G}= h_i^{L}$, where $L$ is the number of layers.

\subsection{Ingredient Set Transformer}
\label{section_transformer}
To better encode the ingredients in a recipe, we propose an ingredient set transformer to incorporate the ingredient interactions. Specifically, given a recipe with \textit{n} ingredients, we have a matrix $Q \in \mathbb{R}^{n \times d}$ of \textit{n} query vectors with dimension $d$. We define an attention function $Att$ maps $Q$ to outputs using $p$ key-value pairs with $K \in \mathbb{R}^{p \times d}, V \in \mathbb{R}^{p \times d}$:
\begin{equation}
Att(Q, K, V) = \omega \left(Q K^\top \right) V,
\end{equation}
where $\omega$ is the scaled activation function (we use $softmax(\cdot/\sqrt{d})$ in our model) and $Att(Q, K, V)$ is the weighted sum of $V$. We further incorporate the multi-head attention by projecting $Q, K, V$ into different subspaces. Considering two ingredient sets $X_1, X_2 \in \mathbb{R}^{n \times d}$, the multi-head attention for $X_1$ on $X_2$ is defined as:
\begin{equation}
Att(X_1, X_2) = \omega [ (X_1 \cdot W_Q)(X_2 \cdot W_K)^\top ] (X_2 \cdot W_V),
\end{equation}
where $W_Q, W_K, W_V\in \mathbb{R}^{d \times d_o}$ are the trainable weights and $d_o$ is the output dimension. 
Then we feed this self-attention into a Multi-head Attention Block (MAB) and a Set Attention Block (SAB) \cite{set_transformer} to get the transformer-based recipe embeddings $h_i^{TF}$ of recipe $v_i$:
\begin{equation}
\begin{aligned}
MAB(X_1, X_2) &= FFN[W_{m} \cdot X_1 + Att(X_1, X_2)],\\
SAB(X) &= MAB(X, X),\\
h_i^{TF} &= FFN[SAB(X)],
\label{eq5}
\end{aligned}
\end{equation}
where $FFN$ is a feed-forward neural network and $W_{m}\in \mathbb{R}^{d \times d_o}$ is a trainable weight. We then combine the generated $h_i^{TF}$  with the GNN-based learned embedding $h_i^{G}$ to obtain the final embedding for each recipe node, denote as $h_i = W_{O} \cdot (h_{i}^{G} + h_{i}^{TF})$, where $W_{O} \in \mathbb{R}^{d_o \times d_o}$ is a weight for output projection. For each user node, we have $h_i =h_{i}^{G}$.

\subsection{Objective Function for Recommendation}
\label{section_objective_function}
We employ a ranking-based objective function to train the model. In particular, the objective involves comparing the preference scores between nodes connected by a \textit{user-recipe} relation against the scores between an arbitrary pair of user and recipe nodes. For example, given a connected user node $v_i$ and a recipe node $v_j$, we encourage the score $s_{i,j}$ between $v_i$ and $v_j$ to be higher than the score $s_{i,j'}$ between $v_i$ and a random negative recipe node $v_j'$. Accordingly, we use a predictor (e.g., inner product) to calculate the score and formulate the recommendation loss $L_{rec}$ as follows:
\begin{equation}
\begin{aligned}
s_{i,j} &= score\_predictor(h_i, h_j),\\
L_{rec} &= \sum_{i \in \mathcal{U}, j\in N_{i}} max(0, 1-s_{i,j}+s_{i,j'}),
\end{aligned}
\end{equation}
where $h_i, h_j, h_j'$ are the embeddings of nodes $v_i, v_j, v_j'$ respectively, $\mathcal{U}$ is the set of users, and $N_{i}$ indicates the recipe neighbors of $v_i$.

\begin{figure*}[t]
	\centering
	\includegraphics[width=\textwidth]{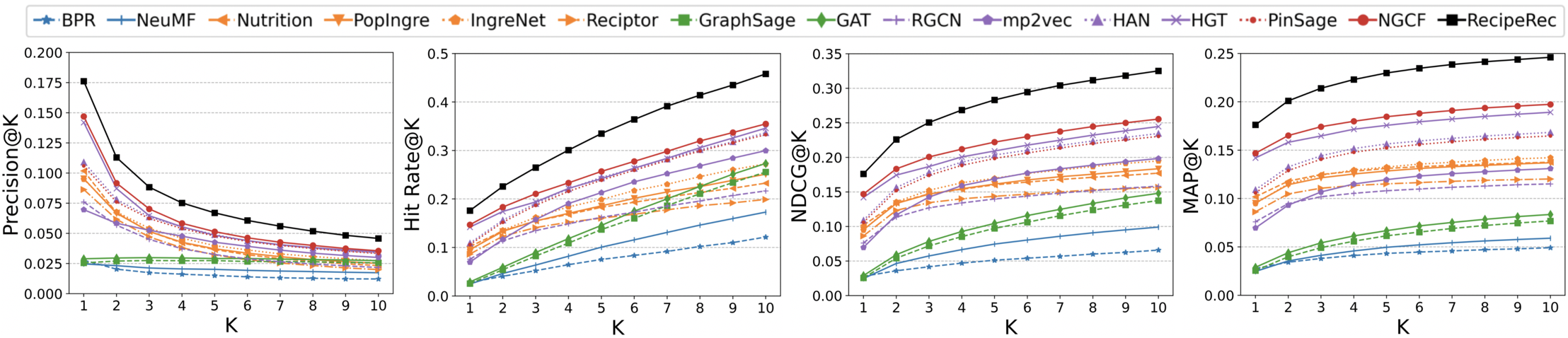}
	\caption{Performance of Top-\textit{K} recipe recommendation where \textit{K} ranges from 1 to 10.}
	\label{fig:result_1}
\end{figure*}

\subsection{Graph Contrastive Augmentation}
Since the above objective function only considers explicit and supervised information in the \textit{URI-Graph}, we further introduce a graph contrastive augmentation strategy to capture graph knowledge in a self-supervised manner. In particular, the input graph $\mathcal{G}$ will be augmented with two correlated views $\hat{\mathcal{G}}', \hat{\mathcal{G}}''$, where $\hat{\mathcal{G}}' \sim q'(\cdot | \mathcal{G}), \hat{\mathcal{G}}'' \sim q''(\cdot | \mathcal{G})$, and $q'(\cdot | \mathcal{G})$, $q''(\cdot | \mathcal{G})$ are two augmentation functions composed of node dropout and edge dropout \cite{ngcf}. We apply these augmentations each training epoch and define a contrastive loss to maximize the agreement between different views of the same node, compared to that of other nodes. Specifically, we treat the views of the same node as positive pairs (i.e., $\{(h_{i}^{'}, h_{i}^{''})|i \in \mathcal{Y}_T\}$) and those of different nodes that appear in the same batch as negative pairs (i.e., $\{(h_{i}^{'}, h_{j}^{''}) | i \in \mathcal{Y}_T, j \in \mathcal{B}_T, i \neq j, \phi_i=\phi_j \}$). Here $\mathcal{Y}_T$ is the set of users and recipes that appear in the training set, and $\mathcal{B}_{T} \subseteq \mathcal{Y}_{T}$ is a random batch of users and recipes in the training set. We use InfoNCE \cite{infoNCE} to model the graph contrastive learning loss $L_{con}$:
\begin{equation}
L_{con}=\sum_{i\in \mathcal{Y}_{T}} -\log \frac{\exp(h_{i}^{'},h_{i}^{''}/\tau)}
{\sum_{j\in \mathcal{B}_{T}}\exp(h_{i}^{'},h_{j}^{''}/\tau)},
\end{equation}
where $\tau$ is the temperature. The final objective function $L$ is defined as the weighted combination of $L_{rec}$ and $L_{con}$: 
\begin{equation}
L = L_{rec} + \lambda L_{con},
\end{equation}
where $\lambda$ is a trade-off weight for balancing two losses.

\section{Experiments}
In this section, we conduct extensive experiments to compare the performances of different models. We also show ablation studies, parameter sensitivity, case studies and embedding visualization to demonstrate the superiority of \textit{RecipeRec}.

\subsection{Experimental Setup}
We employ the classic leave-one-out method to evaluate the model performance \cite{neumf}. During the testing, we randomly sample 100 negative recipes for each user and evaluate the performance using \textit{Precision} (Pre), \textit{Hit Rate} (HR), \textit{Normalized Discounted Cumulative Gain} (NDCG), and \textit{Mean Average Precision} (MAP). We report the performance under top@\textit{K}, where \textit{K} ranges from 1 to 10.

\subsection{Baseline Methods}
We compare our model with 14 baselines, including traditional recommendation approaches \textbf{BPR} \cite{bpr} and \textbf{NeuMF} \cite{neumf}, recipe representation learning and recommendation methods \textbf{Nutrition} \cite{Teng2012}, \textbf{PopIngre} \cite{Teng2012}, \textbf{IngreNet} \cite{Teng2012} and \textbf{Reciptor} \cite{Reciptor}, homogeneous GNNs \textbf{GraphSAGE} \cite{graphsage} and \textbf{GAT} \cite{gat}, heterogeneous GNNs \textbf{RGCN} \cite{rgcn}, \textbf{mp2vec} \cite{metapath2vec}, \textbf{HAN} \cite{han} and \textbf{HGT} \cite{HGT}, and graph-based recommendation methods \textbf{PinSage} \cite{pinsage} and \textbf{NGCF} \cite{ngcf}.

\subsection{Implementation Details}

For the proposed \textit{RecipeRec}, we set the learning rate to 0.005, the number of attention heads to 4, the hidden size to 128, the temperature $\tau$ to 0.07, $\lambda$ to 0.1, node and edge dropout ratio to 0.1, batch size to 1024 and the training epochs to 100.

\subsection{Performance Comparison}
We report the performances of all models in Figure \ref{fig:result_1}. As shown in the figure, \textit{RecipeRec} outperforms all the baselines for all metrics over different \textit{K}s. Specifically, traditional recommendation approaches (i.e., BPR, NeuMF) perform poorly because they neither consider the higher-order collaborative signal nor the associated ingredients. Recipe recommendation approaches (e.g., PopIngre, IngreNet) obtain decent performance after incorporating the relational information, while Reciptor performs poorly since it is designed to learn recipe representations only, which ignores the user information. Homogeneous GNNs (i.e., GraphSage, GAT) behave poorly because of the neglect of node type information. Heterogeneous GNNs (e.g., HAN, HGT) and graph-based recommendation approaches (i.e., PinSage, NGCF) achieve satisfactory results, but fail to encode the influence of different nodes and relations as well as the complex interactions among nodes. Finally, \textit{RecipeRec} achieves the best performance compared to all the baselines, demonstrating the effectiveness of the proposed model.

\subsection{Ablation Study}
\begin{figure}[t]
	\centering
	\includegraphics[width=0.9\columnwidth]{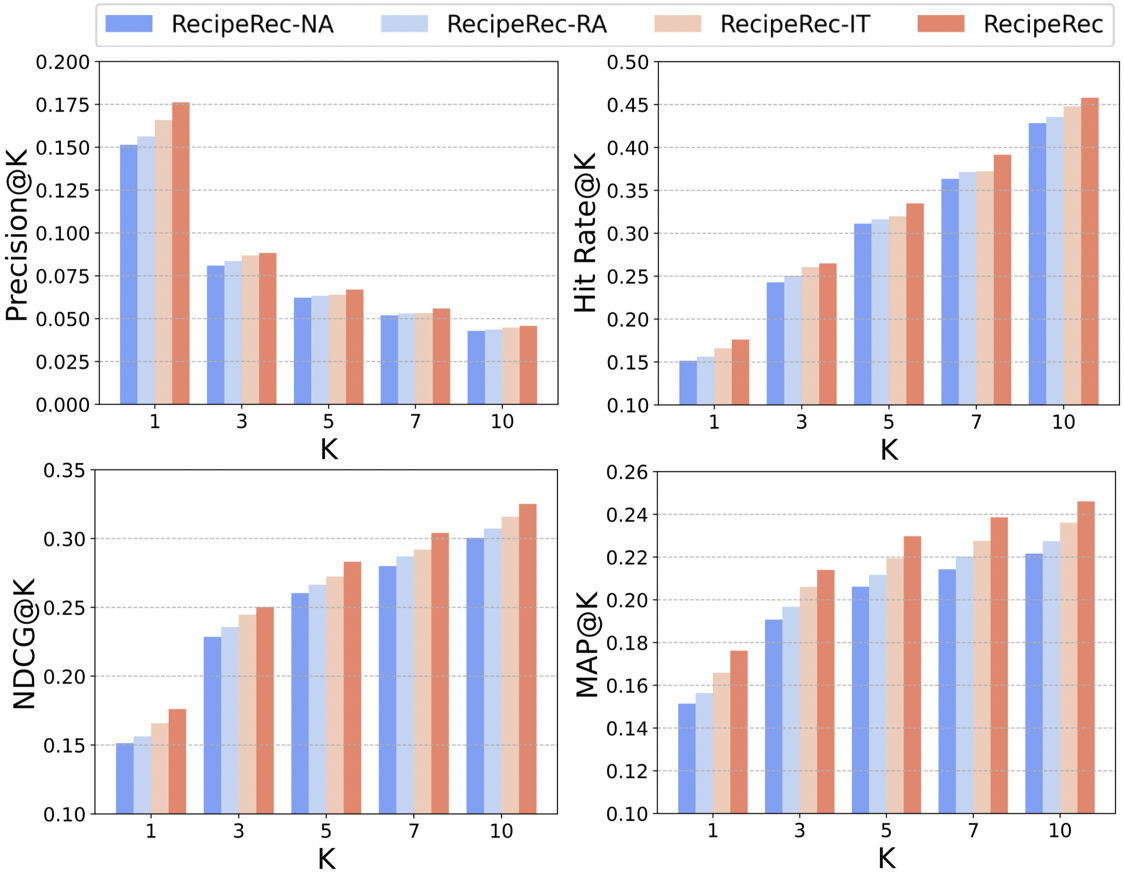}
	\caption{Results of different model variants.}
	\label{fig:ablation_studies_1}
\end{figure}

\textit{RecipeRec} is a joint learning framework composed of several neural network modules and optimized by a joint loss. We further conduct ablation studies to evaluate the performance of different model variants including:
(a) \textbf{\textit{RecipeRec-NA}} that only uses adaptive node-level attention and recommendation loss;
(b) \textbf{\textit{RecipeRec-RA}} that utilizes adaptive node-level attention, relation-level attention and recommendation loss;
(c) \textbf{\textit{RecipeRec-IT}} that combines all neural network modules but only uses recommendation loss.
We report the performances of these model variants under different \textit{K}s in Figure \ref{fig:ablation_studies_1}. 
From the figure, we can find that the performance increases when we incorporate more modules and optimize the model with both losses. Among these variants, the full model \textit{RecipeRec} achieves the best result in all cases. This demonstrates the effectiveness of adaptive node-level attention, relation-level attention, ingredient set transformer and graph contrastive augmentation in our model.

\subsection{Parameter Sensitivity}
We perform parameters sensitivity analyses, including the impact of dropout ratio, trade-off weight, and score predictor.

\noindent
\textbf{Impact of Dropout Ratio in Contrastive Learning.}
We conduct experiments to compare the influence of node dropout and edge dropout in Table \ref{tab:result_dropout_ratio}. We can find that in most cases, the model performs the best when the ratio is 0.1. In general, the performance decreases when further increasing the dropout ratio, which indicates the suitable node/edge dropout ratio will lead to the best model performance.

\noindent
\textbf{Impact of Trade-off Weight in Joint Loss.}
To investigate the influence of graph contrastive learning, we study the impact of trade-off weight $\lambda$ in Table \ref{tab:result_lambda}. We can find that the model achieves the highest performance when $\lambda$ is 0.1. Further increment of the value will decrease the performance. This demonstrates the incorporation of suitable graph contrastive augmentation can lead to the best recommendation results.

\begin{figure}[t]
	\centering
	\includegraphics[width=\columnwidth]{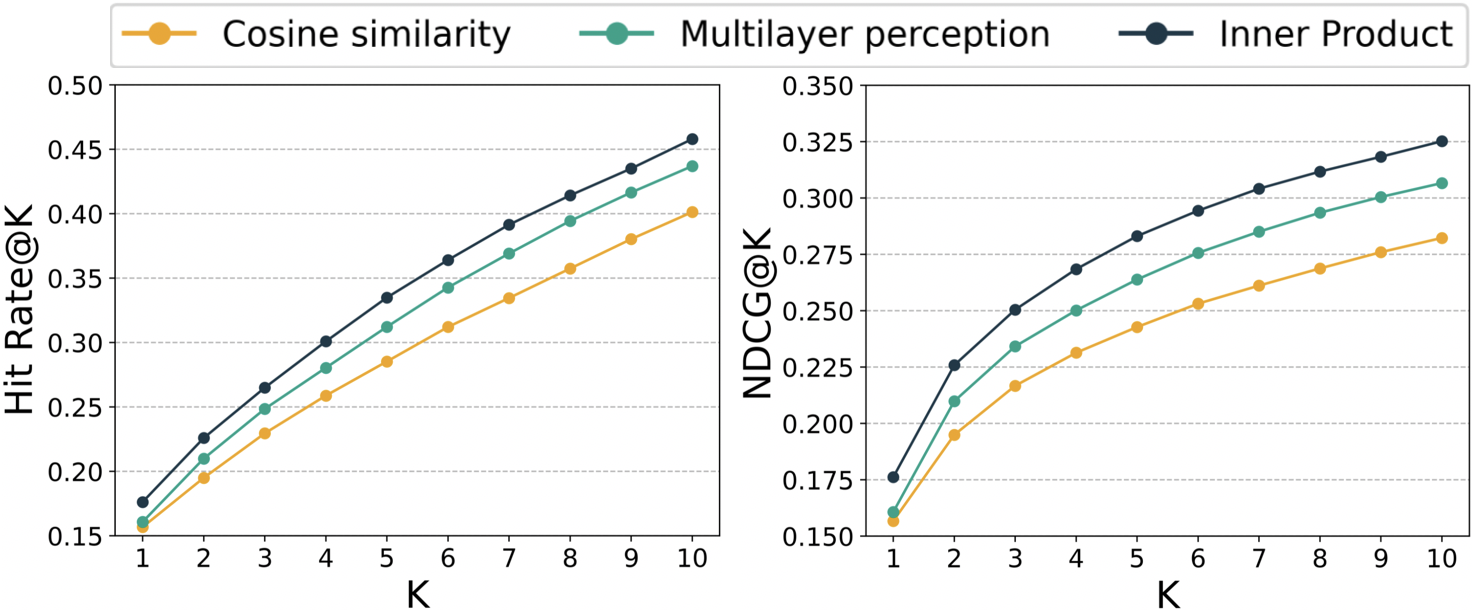}
	\caption{Performance \textit{w.r.t.} the score predictor.}
	\label{fig:result_similarity_func}
\end{figure}

\noindent
\textbf{Impact of Score Predictor.}
\label{section_score_predictor}
To show the effectiveness of the score predictor used in our model, we examine the effect of other predictors (i.e., cosine similarity, multilayer perception) in Fig.\ref{fig:result_similarity_func}. From this figure, we conclude that inner product performs the best since it considers both the angle difference and the magnitude between two embeddings.

\subsection{Case Study: Embedding Visualization}
\begin{figure}[t]
	\centering
	\includegraphics[width=\columnwidth]{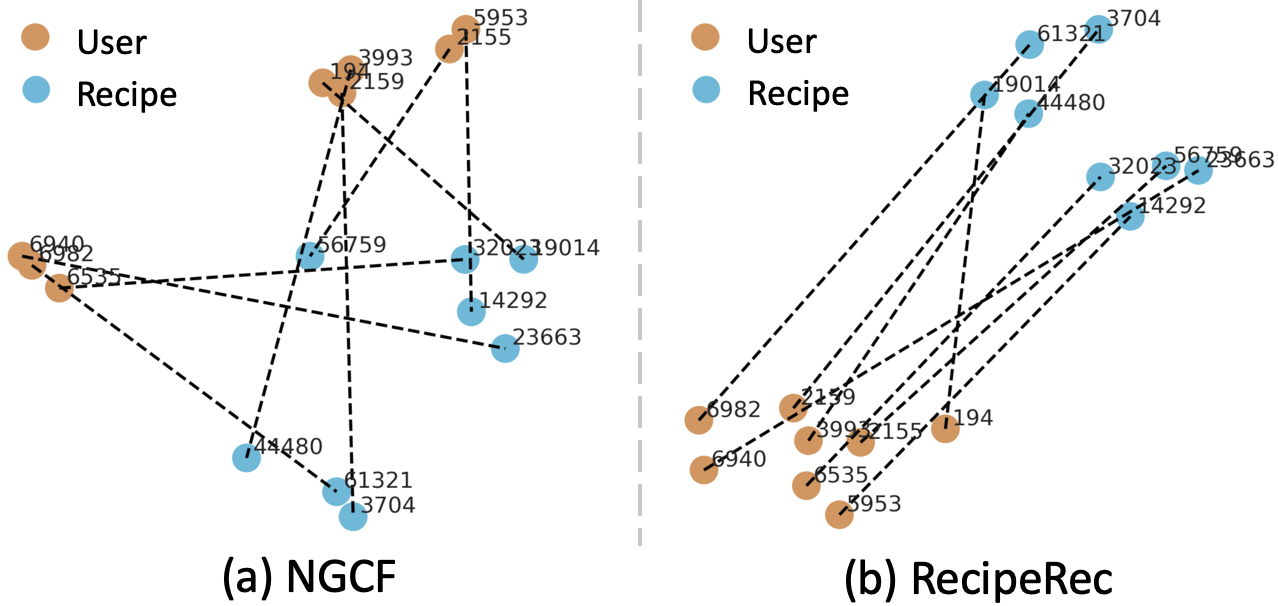}
	\caption{Embedding visualization of user and recipe. The edge indicates the interaction between user and recipe.}
	\label{fig:result_visualization}
\end{figure}

To show a more intuitive understanding and comparison, we randomly select 8 user-recipe interaction pairs and visualize their embeddings using t-SNE. As shown in Fig.\ref{fig:result_visualization}, NGCF can roughly separate the users and recipes into two clusters. There are gaps within each cluster and the lines between clusters are disordered. Ideally, there should be a clear mapping between each user-recipe pair (similar to the `king-man=queen-woman' relationship in \citeauthor{word2vec}~(\citeyear{word2vec})) and the lines should be parallel to each other. However, \textit{RecipeRec} can easily divide the users and recipes into two condense clusters and obtain parallel lines for most pairs, which further demonstrates the superiority of our model.

\begin{table}[t]
\caption{Performance \textit{w.r.t.} dropout ratio in contrastive learning.}
\begin{center}
\resizebox{0.96\columnwidth}{!}{
\begin{NiceTabular}{c|c|cc|cc|cc}
    \toprule
    \multirow{2.5}{*}{Dropout} & \multirow{2.5}{*}{Ratio}& \multicolumn{2}{c}{\textit{K}=1} & \multicolumn{2}{c}{\textit{K}=5} & \multicolumn{2}{c}{\textit{K}=10}\\
    \cmidrule{3-8}
    & & HR & NDCG & HR & NDCG & HR & NDCG \\
    \midrule
    \multirow{5}{*}{Node Drop}
    & \multicolumn{1}{c}{0.1} & 15.49 & 15.49 & \textbf{32.34} & \textbf{27.29} & \textbf{45.13} & \textbf{31.67} \\
    & \multicolumn{1}{c}{0.2} & \textbf{15.77} & \textbf{15.77} & 32.11 & 27.08 & 44.55 & 31.33 \\
    & \multicolumn{1}{c}{0.3} & 15.66 & 15.66 & 31.93 & 27.25 & 44.30 & 31.46 \\
    & \multicolumn{1}{c}{0.4} & 14.99 & 14.99 & 31.69 & 26.51 & 44.79 & 31.01 \\
    & \multicolumn{1}{c}{0.5} & 14.68 & 14.68 & 30.72 & 25.76 & 43.10 & 29.99 \\
    \midrule
    \multirow{5}{*}{Edge Drop}
    & \multicolumn{1}{c}{0.1} & \textbf{17.18} & \textbf{17.18} & \textbf{32.94} & \textbf{27.73} & \textbf{45.12} & \textbf{31.88} \\
    & \multicolumn{1}{c}{0.2} & 17.01 & 17.01 & 32.01 & 27.29 & 44.40 & 31.53 \\
    & \multicolumn{1}{c}{0.3} & 16.10 & 16.10 & 32.54 & 27.49 & 44.43 & 31.55 \\
    & \multicolumn{1}{c}{0.4} & 16.16 & 16.16 & 31.79 & 26.91 & 43.99 & 31.08 \\
    & \multicolumn{1}{c}{0.5} & 16.31 & 16.31 & 31.31 & 26.44 & 43.74 & 30.68 \\
    \bottomrule
\end{NiceTabular}
}
\end{center}
\label{tab:result_dropout_ratio}
\end{table}

\begin{table}[t]
\caption{Performance \textit{w.r.t.} trade-off weight $\lambda$.
}
\begin{center}
\resizebox{0.8\columnwidth}{!}{
\begin{NiceTabular}{c|cc|cc|cc}
    \toprule
    \multirow{2.5}{*}{$\lambda$}& \multicolumn{2}{c}{\textit{K}=1} & \multicolumn{2}{c}{\textit{K}=5} & \multicolumn{2}{c}{\textit{K}=10}\\
    \cmidrule{2-7}
    & HR & NDCG & HR & NDCG & HR & NDCG \\
    \midrule
    \multicolumn{1}{c}{0.1} & \textbf{17.33} & \textbf{17.33} & \textbf{32.88} & \textbf{28.00} & \textbf{45.70} & \textbf{32.37} \\
    \multicolumn{1}{c}{0.2} & 15.79 & 15.79 & 31.64 & 26.74 & 44.87 & 31.27 \\
    \multicolumn{1}{c}{0.3} & 15.96 & 15.96 & 31.95 & 26.88 & 45.11 & 31.38 \\
    \multicolumn{1}{c}{0.4} & 15.73 & 15.73 & 32.41 & 27.19 & 45.33 & 31.60 \\
    \multicolumn{1}{c}{0.5} & 15.39 & 15.39 & 32.50 & 27.10 & 45.06 & 31.38 \\
    \bottomrule
\end{NiceTabular}
}
\end{center}
\label{tab:result_lambda}
\end{table}

\section{Related Work}
This work is closely related to the studies of recipe recommendation and graph neural networks (GNNs). 
\\
\textbf{Recipe Recommendation.}
Existing recipe recommendation approaches are mostly content-based \cite{eating_healthier} or collaborative filtering-based \cite{ensemble_modeling} or hybrid \cite{visually_aware_rec}. However, none of these approaches consider the higher-order collaborative signal such as relations. In our work, we propose a novel heterogeneous graph learning model to encode both recipe content and relational structure information. Additionally, existing recipe recommendation systems usually focus on the rating prediction task, which minimizes the errors between predictions and real ratings. This is concerned with only the observed recipes with explicit ratings, while our work uses a pairwise ranking strategy to examine the recommendation, which accounts for all the recipes in the collection.

\noindent
\textbf{Graph Neural Networks.}
Many graph neural networks \cite{gat,han,hetgnn-kdd19} were proposed to encode the graph-structure data. They take advantage of content information and relation information in graphs to learn node embeddings for various downstream tasks such as recommendation. Recently, several graph-based models have been proposed for recommendation \cite{zhang2017collaborative,pinsage,ngcf}. However, most of them either fail to consider the collaborative signal. In this work, we propose a novel heterogeneous GNN to consider both collaborative signal and recipe information. In addition, contrastive learning has shown great performance in pre-training GNN \cite{gcc}. We thus introduce a graph contrastive augmentation strategy to capture informative graph knowledge and improve the model.

\section{Conclusion}
In this paper, we propose and formalize the problem of \textit{recipe recommendation with graphs}. To solve this problem, we create \textit{URI-Graph}, a new and large-scale recipe graph data to facilitate graph-based food studies. Furthermore, we develop \textit{RecipeRec}, a novel heterogeneous graph learning model for recipe recommendation. \textit{RecipeRec} is able to capture both recipe content and the relational structure information through a graph neural network and a transformer. A novel combined objective function of recommendation loss and graph contrastive learning loss is employed to optimize the model. Extensive experiments demonstrate that \textit{RecipeRec} outperforms state-of-the-art baselines.

\section*{Acknowledgements}
This work is supported by the Agriculture and Food Research Initiative grant no. 2021-67022-33447/project accession no.1024822 from the USDA National Institute of Food and Agriculture.

\bibliographystyle{named}
\bibliography{main}

\end{document}